\documentclass[letterpaper, 10 pt, conference]{ieeeconf}

\IEEEoverridecommandlockouts % This command is only

\usepackage{graphicx} % for pdf, bitmapped graphics files
\usepackage{epsfig} % for postscript graphics files
\usepackage{subfig}
\usepackage{times} % assumes new font selection textcolor{black} installed
\usepackage{amsmath} % assumes amsmath package installed
\usepackage{amssymb} % assumes amsmath package installed
\usepackage{cite}
\usepackage{amsfonts}
\usepackage{bm}
\usepackage{xfrac}
\usepackage{mathdots}
\usepackage{booktabs}
\usepackage{siunitx}
\usepackage{gensymb}
\usepackage{url}
\usepackage{scalefnt}
\usepackage{tikzscale}
\usepackage{tikz}
\usepackage{changepage}
\usetikzlibrary{shapes.geometric}
\usetikzlibrary{calc}
\usetikzlibrary{positioning}
\usetikzlibrary{intersections}

\usepackage{hyperref}
\usepackage{pgfplots}
\usepgfplotslibrary{fillbetween}

\newlength\figureheight
\newlength\figurewidth
\newlength\nodedist

\newlength\fheight
\newlength\fwidth
\title{Real-time Nonlinear MPC Strategy with Full Vehicle Validation for Autonomous Driving}

\author{{Jean Pierre Allamaa$\,\,^\textrm{1}$, Petr Listov$\,\,^\textrm{2}$, Herman Van der Auweraer$\,\,^\textrm{1}$, Colin Jones$\,\,^\textrm{2}$, Tong Duy Son$\,\,^\textrm{1}$}
\thanks{$^\textrm{1}$  Siemens Digital Industries Software,  Leuven, Belgium}
\thanks{Email: \tt \{jean.pierre.allamaa, herman.vanderauweraer, son.tong\}@siemens.com}
\thanks{$^\textrm{2}$  Ecole Polytechnique Federale de Lausanne, Lausanne, Switzerland}
\thanks{Email: \tt \{peter.listov, colin.jones\}@epfl.ch }
\thanks{This work is part of FOCETA project that has received funding from the European Union’s Horizon 2020 research and innovation programme under grant agreement No 956123. The authors also benefited from the ELO-X project under grant agreement No 953348.}
\thanks{Abstract video at: \url{https://youtu.be/tCrn_7331xY}}
}

\begin{document}
	\maketitle
	\thispagestyle{empty}
	\pagestyle{empty}
	
	%%%%%%%%%%%%%%%%%%%%%%%%%%%%%%%%%%%%%%%%%%%%%%%%%%%%%%%%%%%%%%%%%%%%%%%%%
	\begin{abstract}
	In this paper, we present the development and deployment of an embedded optimal control strategy for autonomous driving applications on a Ford Focus road vehicle. Non-linear model predictive control \textcolor{black}{(NMPC)} is designed and deployed on a system with hard real-time constraints. We show the properties of sequential quadratic programming \textcolor{black}{(SQP)} optimization solvers that are suitable for driving tasks. Importantly, the designed algorithms are validated based on a standard automotive XiL development cycle: model-in-the-loop \textcolor{black}{(MiL)} with high fidelity vehicle dynamics, hardware-in-the-loop \textcolor{black}{(HiL)} with vehicle actuation and embedded platform, and full vehicle-hardware-in-the-loop \textcolor{black}{(VeHiL)}. The autonomous driving environment contains both virtual simulation and physical proving ground tracks. NMPC algorithms and optimal control problem formulation are fine-tuned using a deployable C code via code generation compatible with the target embedded toolchains. Finally, the developed systems are applied to autonomous collision avoidance, trajectory tracking, and lane change at high speed on city/highway and low speed at a parking environment.
	\end{abstract}
	%%%%%%%%%%%%%%%%%%%%%%%%%%%%%%%%%%%%%%%%%%%%%%%%%%%%%%%%%%%%%%%%%%%%%%%%%
	\section{Introduction}
	%%%% ADAS Control Applications, NMPC as a viable solution%%%%
	Advanced vehicle control algorithms are crucial for the development of safe and reliable \textcolor{black}{autonomous driving (AD)} applications to reduce road accidents and causalities~\cite{DO2021104856}. Conventional control designs such as PID and linear state feedback often serve in low-level feedback loops of industrial automotive applications. Albeit having low computational complexity, the performance of such controllers is limited in safety-critical traffic scenarios such as emergency collision avoidance. Recently, nonlinear model predictive control for autonomous driving applications has been studied and shown promising results, mainly from academic research~\cite{soncdc2019, 8754713}. On the other side, due to the computational complexity and limited resources required by numerical optimization, NMPC has not been commonly considered in industrial autonomous driving control platforms, \textcolor{black}{with most contributions often omitting one of the fundamental XiL stages~\cite{Ferreau2017}}.

	The main contribution of this paper is an efficient development framework that can be used in the automotive industry to design and safely deploy real-time NMPC on road vehicles. \textcolor{black}{Two test scenarios are created for the intended trajectory tracking application:}
	\begin{itemize}
		\item Autonomous valet parking: to deal with safety around suddenly appearing pedestrians and vehicles. The driving takes place at a low speed, around 10kph. The controller is tested in a private parking area.
		\item Lane keeping: to avoid collision with obstacles in city or highway scenarios. NMPC is tested on proving ground in Aldenhoven, Germany at high speeds around 60kph.
	\end{itemize}
	%%%% NMPC advantages + Literature review %%%%
	We present a framework for embedded NMPC in autonomous driving applications, providing a number of benefits: First, it is built upon a detailed nonlinear predictive model, capturing the system delays, actuator saturation and look-ahead capabilities to generate feasible trajectories that avoid dangerous driving situations. Second, constraints and objectives can be set for specific performance and driver comfort. NMPC is deployed in the loop with the vehicle, without any resampling, as a low-level controller calculating a control policy every iteration respecting hard real-time constraints. Real-time operation implies a dependency on the logical explanation of the solution and most importantly on the numerical optimization scheme execution time~\cite{RTControl}. 
	
	%%%% XiL %%%%

\begin{figure}
		\centering
		\includegraphics[width=0.9\columnwidth]{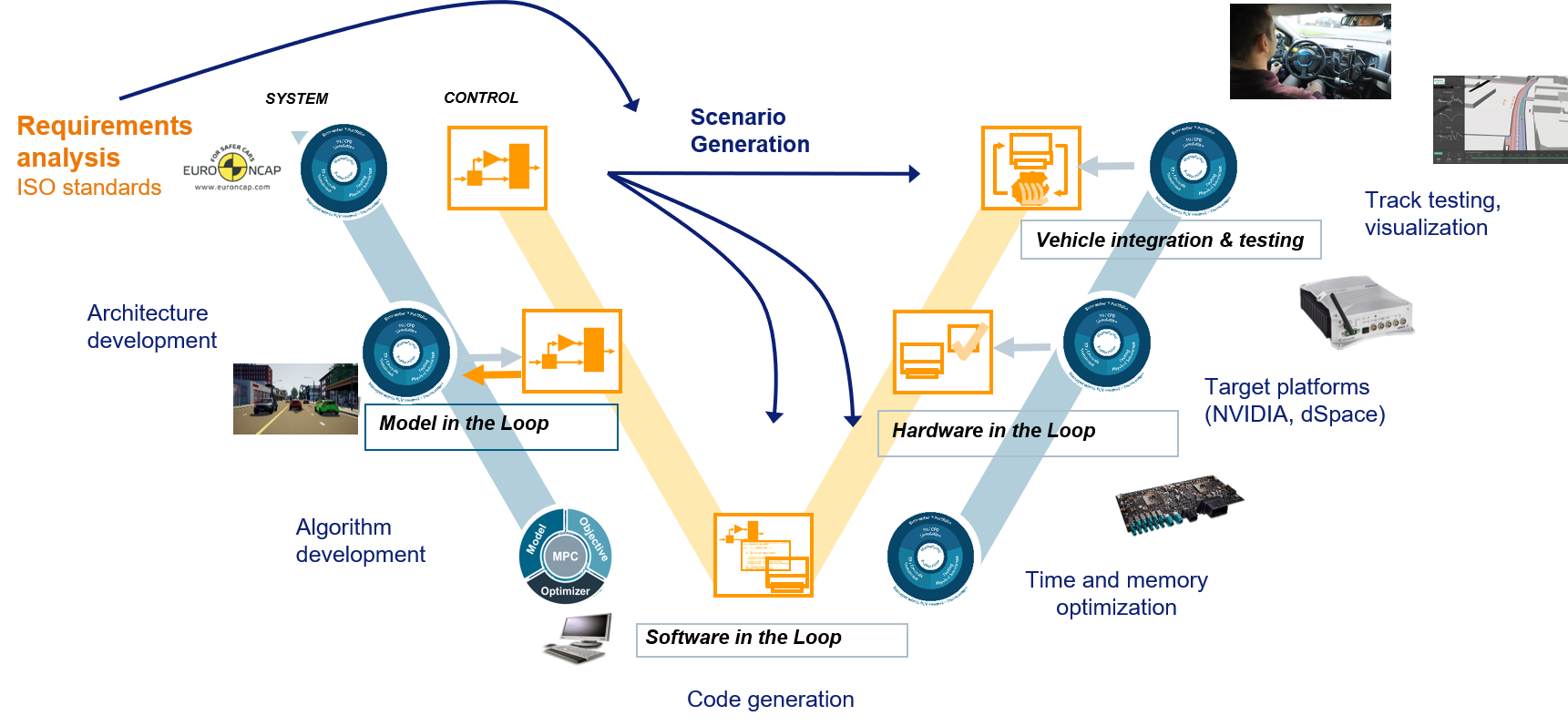}
		\caption{XiL development framework}
		\label{fig:DevelopmentFramework}
	\end{figure}

	The development framework is motivated by the R\&D ADAS team of Siemens, with focus on testing real-time NMPC strategies in different standard scenarios throughout XiL stages~\cite{sontra2017}. XiL verification is well-established as an effective means to develop safe and secure industrial automotive systems, as illustrated in Figure~\ref{fig:DevelopmentFramework}. In this systems engineering process, the control system is first tested in simulation. MiL validation is conducted with high-fidelity multi-physics simulators using Simcenter Amesim with disturbances and parameter mismatches to test the algorithm's robustness. Once validated, C/C++ code is auto-generated and tested (SiL). Real-time performance is validated in a realistic virtual traffic environment with physics-based sensors in Simcenter Prescan. Then the generated code is integrated into ECU hardware (HiL) and eventually deployed in physical road-vehicle (VeHiL). The plug-and-play framework requires limited workforce for the user from design to deployment on the vehicle. Real and virtual environment combined in XiL is attractive, as it helps reducing testing costs as performance is assessed without test sites and costly sensors/obstacles. Tuning campaigns are facilitated, development and implementation cycle time are decreased. \textcolor{black}{This paper presents one demonstration of the NMPC approach applicability in XiL with a decreasing risk factor as we progress through the different stages.} Although tested with a specific toolbox and solver, the framework allows for an ease of reproducibility and expansion to other C/C++ capable toolboxes, solvers,  \textcolor{black}{optimal control problem (OCP)}, car models and scenarios.

	 %%%% high fidelity model for car and sensors (virtual) + actual hardware%%%%
	Hardware setup comprises a Ford Focus vehicle, perception and localization sensors (GPS, Radars, Camera, LiDAR), driving robot (Anthony Best Dynamics) for actuating steering (SR), throttle (AR) and brake systems (BR), and an embedded platform dSPACE MicroAutobox III as the NMPC computation module. They are represented in Figure~\ref{Communication_Confidential}. 
 	
 	The paper is organized as follows. Section II discusses some background on NMPC for autonomous driving. Section III presents the hardware deployment of optimization solvers for collision avoidance application. Validation results with both virtual and real obstacles are given in Section IV.%
	\section{Background}
	%	The objective of this project is to validate the proposed framework by demonstrating a vehicle with Level 3 autonomy for emergency collision avoidance. A single track bicycle model is used since higher DoF models are too complex to handle in embedded applications given the available computational power.	
	This section presents the vehicle model and the developed NMPC formulation for trajectory following. We then provide background on SQP solvers and explain the collision avoidance planning algorithm.

	\subsection{Bicycle Model}
	Car dynamics are represented with a real-time feasible model such as the 6 DoF bicycle model. Controller validation with a 15 DoF model is later performed in a MiL framework. From experimental validation, the model defined below is satisfactory for lane keeping and emergency scenarios, assuming no effect of roll and pitch on lateral dynamics.%
{\small	\begin{align}
			\label{Eq:BicycleModelStates}
				\Dot{v}_x &= \frac{1}{M}(F_{xf} cos\delta + F_{xr} -F_{yf} sin\delta - F_{res} + M\Dot{\psi} v_y), \nonumber\\
				\Dot{v}_y &= \frac{1}{M}(F_{xf} sin\delta + F_{yr} + F_{yf} cos\delta - M\Dot{\psi} v_x), \\
				\Dot{\omega} &= \frac{1}{I_z}(L_f (F_{yf} cos\delta + F_{xf} sin\delta) - L_r F_{yr}), \nonumber\\
	        	\Dot{X} &= v_x cos\psi - v_y sin\psi, \nonumber\\
				\Dot{Y} &= v_x sin\psi + v_y cos\psi, \nonumber\\
				\Dot{\psi} &= \omega. \nonumber
	\end{align}}%
	Linear tire model approximates the lateral forces assuming small slip angles using cornering stiffness $K_f$ and $K_r$. The longitudinal and lateral forces are computed as:%
	{\small	\begin{equation}
			\label{Eq:BicycleModelForces}
			F_{xf} = F_{xr} = 0.5 \frac{t_r T_{max}}{R}, F_{yf} = K_f \alpha_f, F_{yr} = K_r \alpha_r.
	\end{equation}}%
	The front and rear slip angles can be defined as follows:%
	{\small	\begin{equation}
			\label{Eq:BicycleModelSlipAngle}
				\alpha_f = -\tan^{-1}\Big(\frac{\Dot{\psi}L_f + v_y}{v_x}\Big) + \delta,
				\alpha_r = \tan^{-1}\Big(\frac{\Dot{\psi}L_r - v_y}{v_x} \Big) . 
	\end{equation}}%
	\textcolor{black}{Resistance in the longitudinal direction is modeled as the sum of rolling resistance and air drag:}%
	{ \begin{equation}
			\label{Eq:BicycleModelDrag}
			F_{res} = C_{r0} + C_{r2}v_x^2.
	\end{equation}}%
	\noindent\textcolor{black}{The body reference frame has its origin at the CoG with the X-axis pointing to the front of the car. The position of this frame is defined by X,Y in the global frame. We use the body frame to express forces and moments in the car dynamics because the inertia remains constant in it.	Therefore, the kinematics ODE in $X, Y, \psi$ are expressed in the Cartesian global frame as in~\eqref{Eq:BicycleModelStates}. All the variables and model parameters are summarized in Table~\ref{table:BicycleModel_Variables}.}
	
{	\begin{table}
		\centering
		\scriptsize
	\textcolor{black}{	\begin{tabular}{c c c } 
			\hline
			State \& Parameter & Description & Unit \\ [0.5ex] 
			\hline
			$v_x$, $v_y$ & Body frame longitudinal and lateral velocities & \si{\meter}$\cdot$\si{\per\second}\\
			\hline
			$\omega$ & Body frame yaw rate & \si{\radian}$\cdot$\si{\per\second}\\
			\hline
			X,Y & Global Cartesian coordinates & \si{\meter}\\
			\hline
			$\psi$ & Global vehicle heading (yaw) & \si{\radian}\\
			\hline
			M, $I_z$ & Total mass and Inertia& \si{\kilogram}, \si{\kilogram}$\cdot$\si{\square\meter}\\
			\hline
			$L_f$,$L_r$ & CoG's distance from front and rear axles  & \si{\meter}\\
			\hline
			$F_{xf}$,$F_{xr}$ & Local front and rear axles longitudinal forces  & \si{\newton}\\
			\hline
			$\delta$, $t_r$& Steering angle and normalized throttle  & \si{\radian}, 1 \\
			\hline
			$T_{max}$& Maximum engine torque  & \si{\newton}$\cdot$\si{\meter}\\
			\hline
			R& Wheel radius  & \si{\meter}\\
			\hline
			$C_{r0}$, $C_{r2}$ & Zero and second order friction parameters & \si{\newton}, \si{\kilogram}$\cdot$\si{\per\meter}\\
			 \hline
		\end{tabular}}
		\caption{Bicycle model states and parameters}
		\label{table:BicycleModel_Variables}
	\end{table}
}
	\subsection{Nonlinear Model Predictive Control}
	NMPC directly controls the car by computing the normalized throttle $t_r$ with respect to the maximum engine force, and the body frame steering angle $\delta$. A receding horizon scheme of $N$ steps is used. \textcolor{black}{A set of nonlinear difference equations $f_d$ is obtained by applying a Runge-Kutta $4^{th}$ order method to the dynamics $\Dot{x} = f(x,u)$ in~\eqref{Eq:BicycleModelStates}.} The NLP then optimizes over the discrete-time OCP for trajectory following:%
	\textcolor{black}{{\small 	\begin{align}
		\label{Eq:NMPC_Formulation}
		\begin{split}
			\min_{x(0),\dots, x(N),u(0),\dots, u(N-1)} \sum_{k=0}^{N-1}  l_k(x_{k},u_{k}) &+ V_{N}(x_{N})\\
			\textrm{subject to: }  x_{0} = x(0) \\
			x(k+1) = f_d(x(k),u(k)) \qquad &k=0,\dots,N \\
			v_{x,min} \leq v_x(k) \leq v_{x,max} \qquad &k=0,\dots,N-1\\
			v_{y,min} \leq v_y(k) \leq v_{y,max} \qquad &k=0,\dots,N-1\\
			\omega_{min} \leq \omega(k) \leq \omega_{max}  \qquad &k=0,\dots,N-1\\
			X,Y \in \mathcal{D}  \qquad &k=0,\dots,N-1\\
			\delta_{min} \leq \delta \leq \delta_{max} \qquad &k=0,\dots,N-1\\
			-1 \leq t_r \leq 1 \qquad &k=0,\dots,N-1\\
			e_N \in \chi_N. \qquad & \\
		\end{split}
	\end{align}}}%
	The stage cost $l_k(x_{k},u_{k})$ with \textcolor{black}{$u = [\delta, t_r]$} is defined as:%
	{\small	\begin{align}
				l_k(x_k, u_k) = (x(k) - x_{ref}(k)) ^ T Q (x(k) - x_{ref}(k))\\
				\;\;\; +\; u(k) ^ T R u(k) + \Delta u(k) ^ T S\Delta u(k), \nonumber
		\end{align}}%
	 with $Q \in\mathbb{R}^{6\times 6} \succeq 0, R \in\mathbb{R}^{2\times 2} \succ 0, S\in\mathbb{R}^{2\times 2} \succeq 0,$ and $x_{ref}(k) = [v_{ref}(k), 0,0,X_{ref}(k), Y_{ref}(k),\psi_{ref}(k) ]$ is the reference trajectory from the planner. \textcolor{black}{The lateral velocity and yaw rate are regulated to a zero reference. In this application, constraints on the input rates are relaxed and added to the cost function by penalizing large control temporal differences $\Delta u = u(k+1) - u(k) $. $V_N(x_N)$ is a quadratic terminal cost on the tracking error with a bigger cost matrix than Q. $\chi_f$ is the terminal set on the tracking error $e_N = x(N) - x_{ref}(N)$.}
	\textcolor{black}{Moreover, $\mathcal{D}$ is the set of left and right boundaries that define a safe driving corridor for obstacle avoidance as seen in Figure~\ref{fig:Planner}.
%		 NMPC is tasked to drive within this area to guarantee safe collision-avoidance.
}
	\subsection{Collision avoidance planner}
\begin{figure}
		\centering
		\includegraphics[width=0.65\columnwidth]{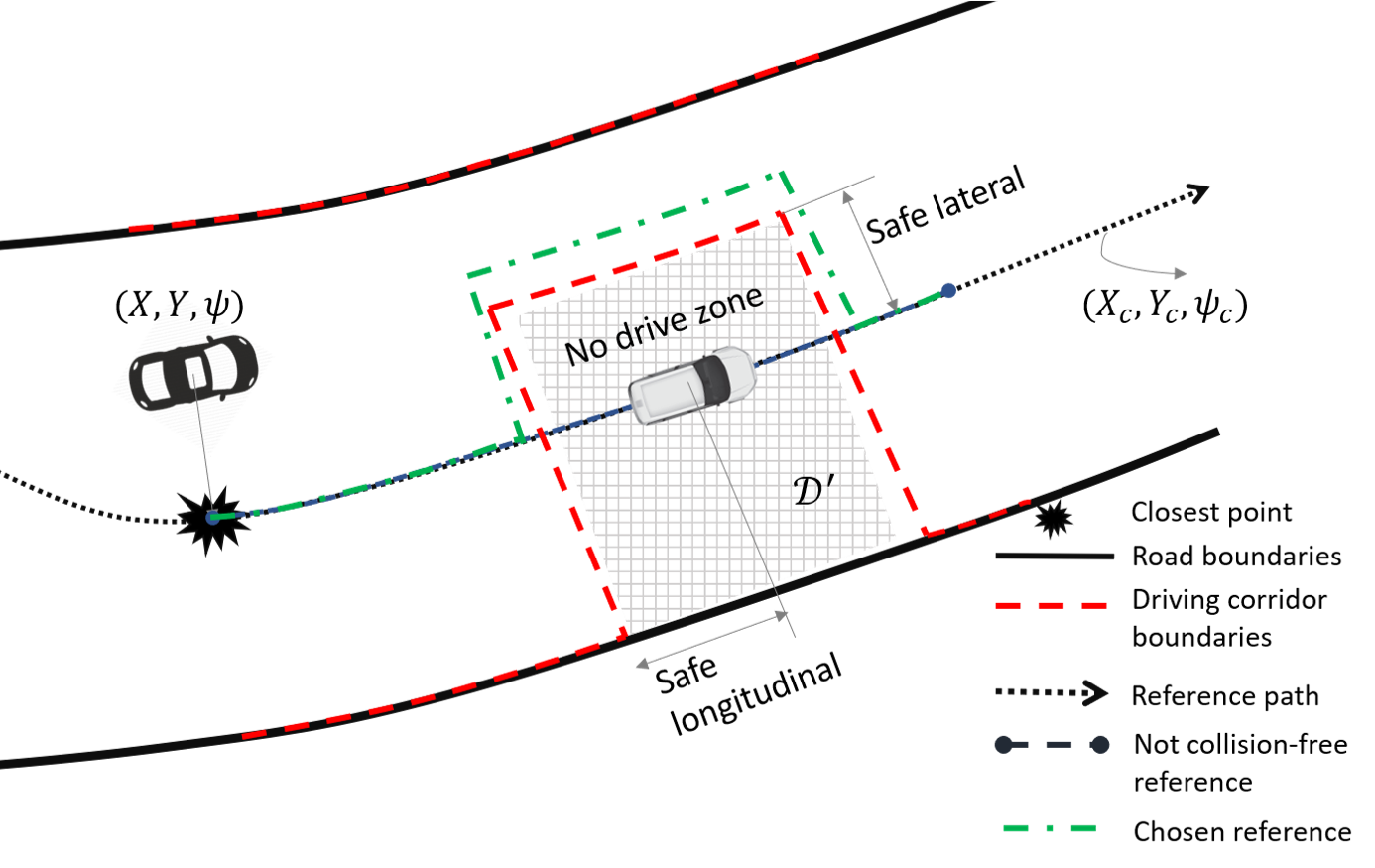}	
		\caption{\textcolor{black}{Planner: closest point localizer and safe corridor}}
		\label{fig:Planner}
	\end{figure}

	\textcolor{black}{The local planner is provided with the original reference path and velocity profile $v_{x,ref}$. It chooses a portion of the global path to track $x_{ref}(k) =[v_{x,ref}(k), 0, 0, X_c(k), Y_c(k), \psi_c(k)]$ for $k \in [k^*,\dots,k^* + N]$ by localizing the car with respect to the closest point on the trajectory such that: $k^* = \underset{k}{\arg\min} (X-X_c(k))^2 + (Y - Y_c(k))^2$. Moreover, the planner receives obstacle information and adjusts the boundaries of the driveable area $\mathcal{D}$ creating a driving corridor taking all obstacles into account as seen in Figure~\ref{fig:Planner}, all while adjusting the reference to be within $\mathcal{D}$. The lateral and longitudinal safe distances create a no-go box-zone $\mathcal{D}^{'}$ around the different obstacles and are tunable online to mimic different reaction times or distances. The size of $\mathcal{D}$ depends on the speed and reaction time. For scenarios presented in this paper, 1.2 to 1.5 seconds of safe duration proved to be sufficient to react to sudden obstacles. }
	\subsection{SQP and QP solvers}
	Sequential quadratic programming is a popular optimization 
	method thanks to its ability to handle highly nonlinear problems and to efficiently warm-start~\cite{NoceWrig06}. In this application, an SQP with QRQP quadratic solver from CasADi is used~\cite{NLPCodeGen}. QRQP is based on the sparsity exploiting active-set method and the derivatives are produced by the CasADi toolbox~\cite{Andersson2019}, an open source symbolic software framework for nonlinear programming.
	This paper does not provide a benchmark/comparison of different solvers for automotive applications but rather implements an out-of-the-box solver to demonstrate the framework. The choice of toolbox and solver is just a placeholder for any other library written in C/C++ or with C code generation capabilities~\cite{verschueren2020acados, PolyMPC}.%
	\section{Embedded NMPC Implementation}
\begin{figure}
		\centering
		\includegraphics[width=0.7\columnwidth]{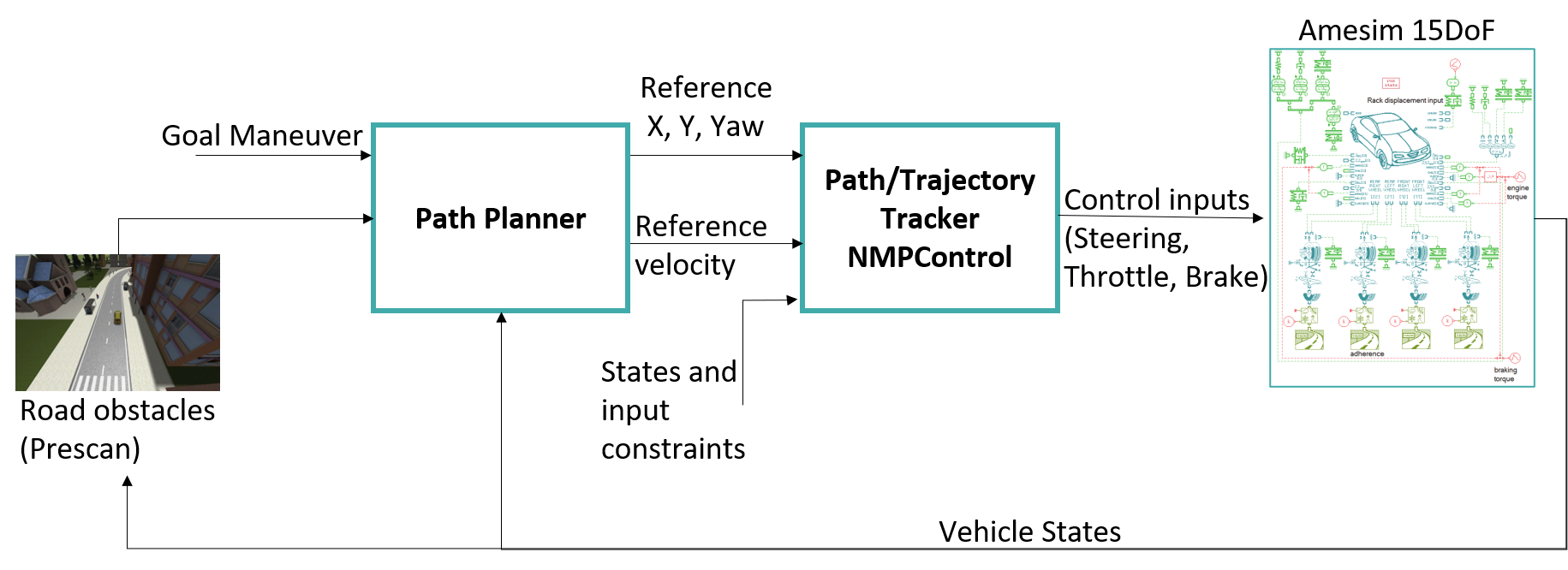}	
		\caption{MiL: closed-loop control structure}
		\label{fig:MiLBlockDiagram}
	\end{figure}

	This section deals with the deployment steps on dSPACE MicroAutobox III (MABX-III) hardware for an embedded control application and validates MiL/HiL in the virtual environment of Prescan. First, the controller is prepared for a real-time environment on a platform with a C/C++ compiler: MABX-III has an ARM Cortex A-15 processor, operates on a 2GB DDR4 RAM with 64MB flash memory for real-time application, and runs RTOS. Deployment to MABX-III is done via Simulink interface. This chapter explains how the tailored NMPC is C-code generated, tested in Simulink, and deployed as a standalone library in runtime. Turning the optimization function into C-code enhances performance with no callbacks into Matlab environment, and with static memory allocation of the block's internal states on compile time. Second, the generated code is validated in MiL simulation with Amesim as in Figure~\ref{fig:MiLBlockDiagram}. 
	\subsection{Code generation for standalone NMPC}
	\begin{figure}[b]
		\centering
		\includegraphics[width=8cm]{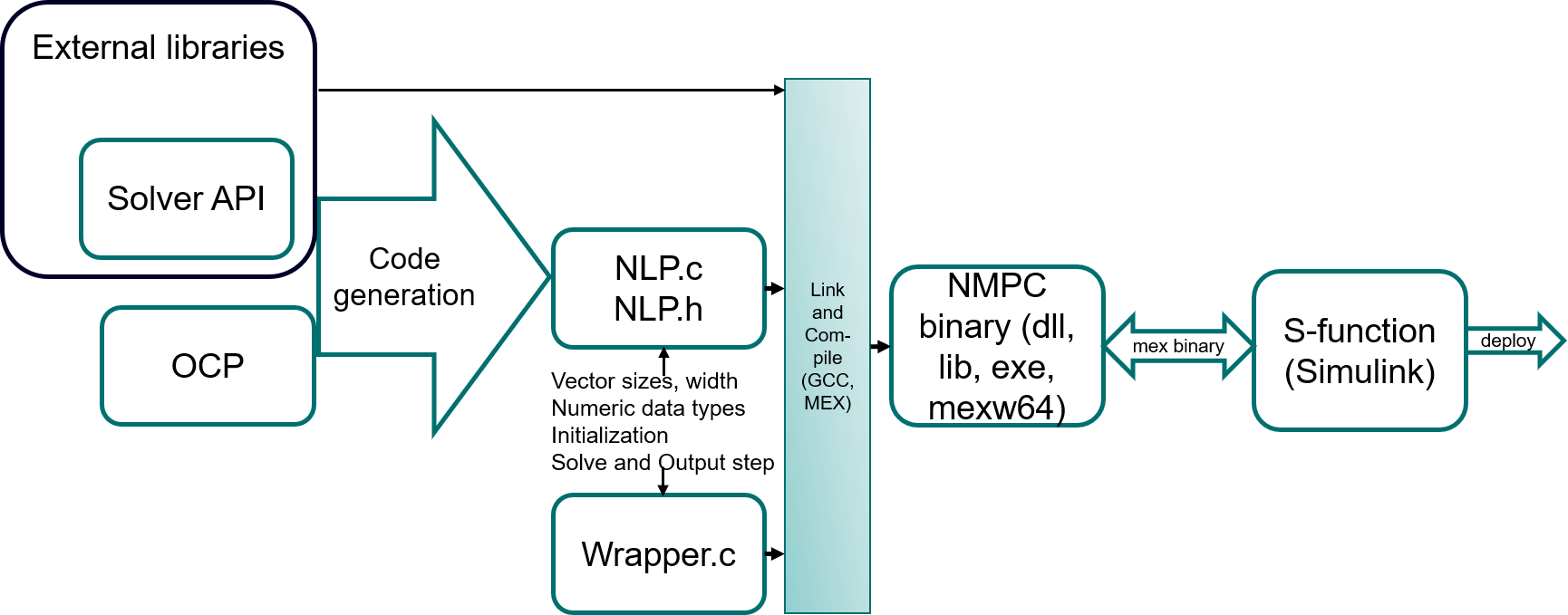}
		\caption{HiL: NMPC deployment steps}
		\label{fig:StepsForDeplyoment}
	\end{figure}
	NMPC for trajectory following is written in Matlab using CasADi. The challenge resides in code generating the Simulink model to be deployed on MABX-III and the proposed procedure is depicted in Figure~\ref{fig:StepsForDeplyoment}. Particularly for this application, the C source codes are compiled using MEX (with MSVC: Microsoft Visual Studio or MinGW64 compilers). Other compilers such as GCC could be used depending on the target platform. For an OCP written in C/C++, the code can be called directly using \textit{S-function builder}. The MEX executable is compiled from the source codes, linked with the toolbox's implementation code, and is called using an \textit{S-Function block}. Therefore, the MEX binary is a self-contained library. The following comments are to be stated for this project implementation: \textcolor{black}{1) CasADi code generation toolbox is used to generate the NMPC source code containing the evaluation of the different steps in an SQP algorithm and construction of the QP subproblems 2) The wrapper for the NMPC source code solves~\eqref{Eq:NMPC_Formulation} given the current state estimates and reference, and outputs an open-loop primal solution and the first control action.}
	
	\subsection{MiL in a virtual environment}
\begin{figure}% 
		\centering
		\includegraphics[width=0.7\columnwidth]{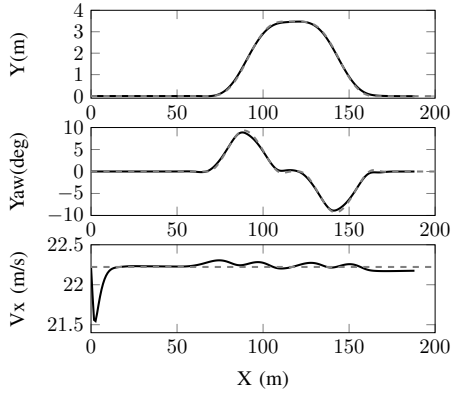}
		\caption{MiL: vehicle states (solid) and reference (dashed)}
		\label{fig:amesim_ISODLC_XYYaw}
	\end{figure}

	The first test in the XiL process, evaluates the closed-loop performance, with high fidelity vehicle dynamics (15DoF) from Amesim, with noise and parameter mismatch. Simulations are carried out to test the NMPC in trajectory following and emergency collision avoidance applications in an ISO 3888-1 standard double lane change scenario at 80kph. \textcolor{black}{Results presented in Figure~\ref{fig:amesim_ISODLC_XYYaw}, show a high-performance tracking response.} NMPC is solved to convergence and satisfies the real-time constraints. The system profiler (Simulink or MSVC profiler), indicates that the code generated NMPC execution time averages at 2.4ms. The effect of code generation on NMPC's execution time is presented in Table~\ref{table:Profiling}. Since MABX-III has limited computational power compared to the host PC (with 32GB RAM), \textcolor{black}{it is necessary to understand how execution time scales on the platform for real-time performance on MABX-III}. After running several HiL tests, it is found that execution time is increased between 7 and 10 times on MABX-III. Therefore, the first validation for real-time performance before deployment is to operate the NMPC on the host PC with an execution time below 4.5ms for a sample time of 40ms. Otherwise, NMPC is real-time incapable and needs to be redesigned. %
	{ \begin{table}[b]
		\centering
		\scriptsize
		\begin{tabular}{|m{2.3cm} | m{2.1cm}| m{2.3cm}  |} 
			\hline
			Type: & No code generation & Code generation\newline
			 (per evaluation) \\ [0.5ex] 
			\hline\hline
			Total & 22.0ms & 2.4ms (2.4ms) \\
			\hline\hline
			QP & 2.00ms & 0.218ms (0.109ms)\\ 
			\hline
			Line search & 2.00ms & 0.218ms (0.109ms)\\
			\hline
			Cost and constraints  & 2.00ms & 0.218ms (0.109ms)\\
			\hline
			Gradient  & 1.00ms & 0.109ms (0.109ms)\\
			\hline
			Hessian  & 8.00ms & 0.87ms (0.436ms)\\
			\hline
			Jacobian  & 5.00ms & 0.545ms (0.27ms)\\
			\hline
		\end{tabular}
		\caption{Profiling optimization sub-functions' evaluation time with and without code generation on Host PC}
		\label{table:Profiling}
	\end{table}}%
\subsection{HiL in a virtual environment}
HiL validation is performed for the same scenario as in MiL, however, controls are applied to a vehicle physically lifted off the ground. Controller real-time capabilities and control signal smoothness are the main targets. Two important aspects are required before deployment: a. all Simulink blocks are code generatable, b. NMPC formulation is code optimized for the quickest execution time. Code generating the optimization function speeds up the evaluation time from 4 to 10 times as compared to the Matlab evaluation~\cite{Andersson2019}. The following properties are used in the NMPC compilation:%
\begin{itemize}
	\item Solver: SQP method with QRQP (Active-Set method)
	\item Maximum number of SQP: 50 and QP: 100 iterations
	\item Integration type: Runge-Kutta 4 (RK4) with 4 steps
	\item Hessian approximation: Exact
	\item OCP method: Direct (discretize then optimize)
	\item Shooting method: Multiple shooting
	\item Sample time: 40ms (25Hz controller)
	\item Prediction horizon: N = 30
	\item Primal and dual infeasibility threshold: $1e^{-06}$ and $1e^{-04}$ 
\end{itemize}%
Possible improvements to reduce computation time are:
\begin{enumerate}%[label=\arabic*)]
	\item Warm start as SQP is heavily affected by initialization
	\item Reformulate the OCP to relax non critical active constraints, add slack variables or add them to the cost
	\item Scale the problem in order to improve conditioning
	\item Reformulate the OCP in a matrix form and minimize nested for-loops to facilitate derivative calculations
\end{enumerate}
The second test towards full vehicle deployment is a validation of the NMPC on MABX-III for highway and cut-in scenarios. The car is visualized in Prescan through communication via ROS~\cite{288} as in Figure~\ref{Communication_Confidential}. \textcolor{black}{NMPC operates as the low-level controller running at 25Hz as the optimal policy is applied on the driving robot. HiL test is validated with the NMPC block operating in real-time on MABX-III, with an execution time around 22ms.}
\section{Vehicle Hardware in the Loop (VeHiL)}
\begin{figure}%
	\centering
	\includegraphics[width=0.7\columnwidth]{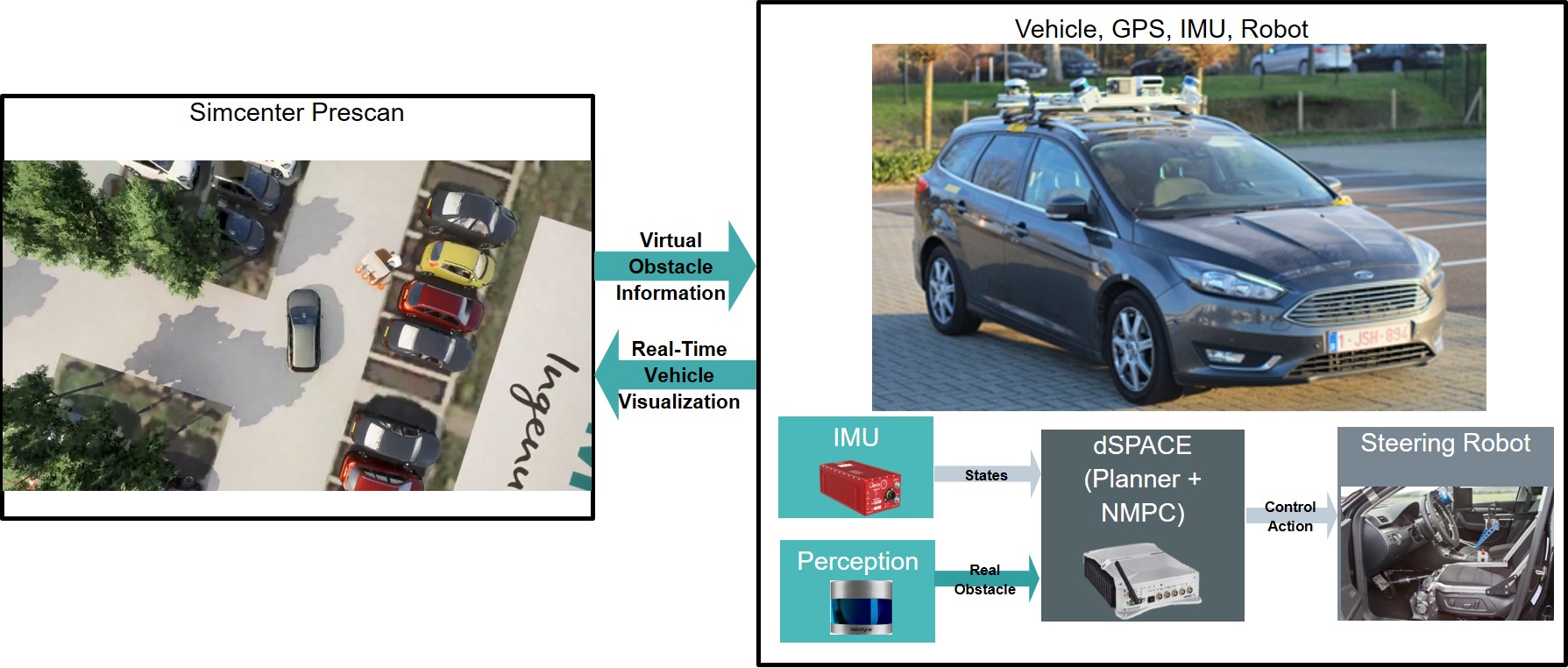}	
	\caption{VeHiL: communication and vehicle framework}
	\label{Communication_Confidential}
\end{figure}

Real-time optimal control implementation for HiL and VeHiL requires communication among hardware and software, presented in this chapter and summarized in Figure~\ref{Communication_Confidential}. \textcolor{black}{The results of the physical testing campaigns in a private parking area and on proving ground are also presented.} 
For HiL testing, the vehicle in Figure~\ref{Communication_Confidential} is replaced by a simulator. Therefore, the communication architecture allows the user to go from offline simulation to online MiL/HiL and finally to VeHiL with the exact same NMPC code.
\subsection{VeHiL: Parking validation}
\begin{figure}%
	\centering
	\includegraphics[width=0.7\columnwidth]{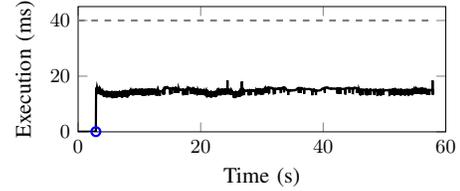}
	\caption{Parking VeHiL: NMPC execution time (solid) and sample time (dashed)}
	\label{fig:Testing_ExecutionTime}
\end{figure}%
\begin{figure}%Uncomment
	\centering
	\includegraphics[width=0.7\columnwidth]{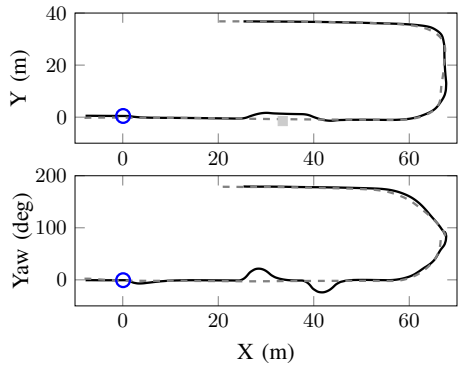}
	\caption{Parking VeHiL: X-Y and X-Yaw states (vehicle states: solid, reference trajectory: dashed, Obstacle: box)}
	\label{fig:Testing_XY}
\end{figure}%
\begin{figure}%Uncomment
	\centering
	\includegraphics[width=0.7\columnwidth]{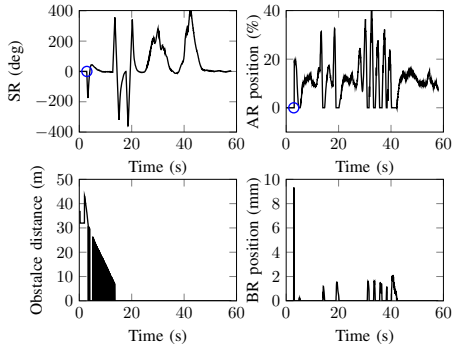}	
	\caption{\textcolor{black}{Parking VeHiL: obstacle and NMPC commands}}
	\label{fig:Testing_ObstacleDetection}
\end{figure}%
In this testing campaign, the vehicle is integrated in the loop with the embedded controller for parking scenarios, in presence of obstacles. The original reference trajectory is generated from human driving around the parking area at 10kph, without obstacles. MiL with Amesim and HiL with the virtual environment are first validated for this scenario. \textcolor{black}{Although safe, autonomous driving was uncomfortable with jerky throttle and aggressive steering. This could be caused by the stalling engine torque at low speeds and the layout of the parking area that included an upward slope, which were not accounted for in the NMPC dynamics.} Figure~\ref{Communication_Confidential} shows a real-time visualization of the physical Ego car generated by colleagues at the ADAS group in Siemens: \textcolor{black}{MABX-III receives the vehicle coordinates from IMU and GPS and communicates with Prescan for a virtual representation of the testing site. Moreover, Ego vehicle controlled by NMPC can be seen in grey avoiding a virtual construction person.}

Figure~\ref{fig:Testing_XY} shows the autonomously driven path after the controller tuning campaign for this AD scenario. NMPC commands the robot as the $X=0m$ line is crossed (blue circle). Those plots demonstrate the NMPC capabilities in both accurate tracking and collision avoidance. The planner shifts the reference laterally from the original one for a lane change at 1.5m. NMPC reacts quickly to an obstacle only detected within 10m of distance. Finally, Figure~\ref{fig:Testing_ObstacleDetection} shows the obstacle detected range and the NMPC commands in steering robot angle (SR), throttle pedal position (AR), and brake (BR). The obstacle information is not fed to the NMPC before a distance-to-collision of 10 meters, to simulate emergency obstacle avoidance. \textcolor{black}{The unsmooth throttle behavior between 30 and 40s is caused by the vehicle deceleration in the upward slope. This can be tackled by tuning, a more accurate model, or by online parameter adaptation. Nevertheless, the task was still accomplished with real-time performance as the embedded NMPC solves with an execution time of 14ms for a horizon of 1.2seconds as shown in Figure~\ref{fig:Testing_ExecutionTime}.}
\subsection{VeHiL: Proving ground validation}
\begin{figure}
	\centering
	\includegraphics[height = 4cm,width=0.9\columnwidth]{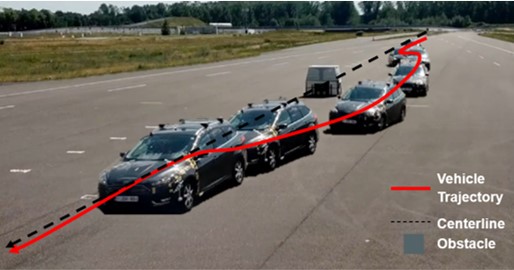}	
	\caption{Proving ground VeHiL: obstacle avoidance}
	\label{fig:dlc_andenhoven}
\end{figure}
\begin{figure}%Uncomment
	\centering
	\includegraphics[width=0.7\columnwidth]{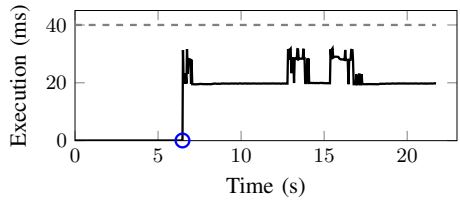}	
	\caption{Proving ground VeHiL: NMPC execution time (solid) and sample time (dashed)}
	\label{fig:AldenDay2_ExecutionTime}
\end{figure}
\begin{figure}
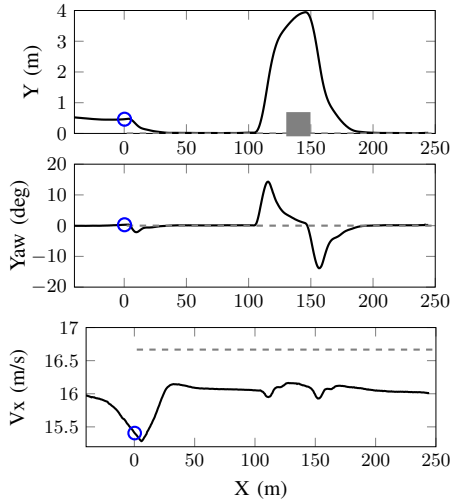
%
	\centering
	\includegraphics[width=0.7\columnwidth]{Figures/12_XYYaw_Pos_2.tikz}	
	\includegraphics[width=0.7\columnwidth]{Figures/12_Vx_2.tikz}	
	\caption{Proving ground VeHiL: vehicle X-Y,  X-Yaw, longitudinal speed (vehicle states: solid, reference trajectory: dashed, NMPC activation: blue, Obstacle: box)}
	\label{fig:AldenDay2_XYYaw}
\end{figure}
\begin{figure}%Uncomment
	\centering
	\includegraphics[width=0.7\columnwidth]{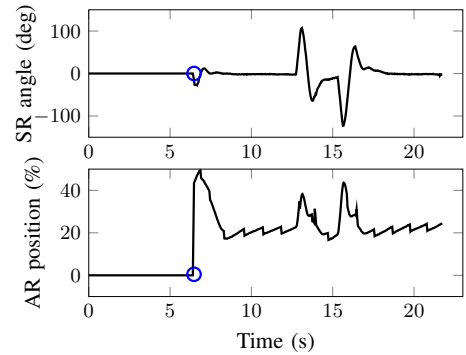}	
	\caption{Proving ground VeHiL: NMPC control actions}
	\label{fig:NMPCControl_Aldenhoven}
\end{figure}

The second part of the campaign took place in Germany on a secure testing site as shown in the shots of Figure~\ref{fig:dlc_andenhoven} and consisted of an \textcolor{black}{Adaptive Cruise Control (ACC)} for lane-keeping at 60kph over 500 meters with collision avoidance using a dummy vehicle. \textcolor{black}{In order to test the NMPC at its boundaries of sudden obstacle emergence, the safe duration parameter, presented in Figure~\ref{fig:Planner} was reduced online from 1.8 to 1.2 seconds of reaction time (20 meters of safe distance before and after the obstacle at 60kph).} This step, shows the benefits of deploying such a controller in automotive applications as the duration is insufficient for the human driver to take control and avoid an accident. Results of this testing phase, shown in Figures~\ref{fig:AldenDay2_ExecutionTime} through \ref{fig:NMPCControl_Aldenhoven}, prove the NMPC quickly reacted while satisfying all dynamic and actuator constraints. The controller smoothly corrects the initial positioning error off the centerline and performs a cruise control until obstacle detection. The car immediately recovers the original lane after avoiding the obstacle. NMPC's execution time averaged at 22ms, for a sampling time of 40ms, hence running in real-time on the embedded platform as in Figure~\ref{fig:AldenDay2_ExecutionTime}. The velocity tracking error is similar to the first phase, at almost 0.5m/s as in Figure~\ref{fig:AldenDay2_XYYaw} and is mainly due to longitudinal model mismatches.
As from Figures~\ref{fig:Testing_ObstacleDetection} and \ref{fig:NMPCControl_Aldenhoven}, the jerky throttle control could be a result of the robot delay and the step change in the spatial reference causing some constraints to become active.

Numerical convergence was achieved within at most 2 SQP and 1 to 3 QP iterations, for a total computation time of 22ms and 14ms on average for the 60kph and 10kph scenarios respectively. \textcolor{black}{Warm starting the primal variables, taking into consideration the driving corridor, significantly cut down execution time.} The SQP solver used in this project scenarios was efficient and satisfactory in real-time optimal control, handling non-linearities and converging to the optimal solution. The scenarios were carried out in a MiL framework and resulted in control policies similar to VeHiL, creating a potential real to simulation flow. This shows the developed framework's importance as most of the tuning, OCP reformulation, and real-time capabilities were validated with the high fidelity model, with little costs, zero incidents, and safe vehicle integration. \textcolor{black}{XiL cycle testing improves scalability and sensitivities to parameters as it allows testing with various driving scenarios and traffic situations all while injecting disturbances and noises in simulation and hardware.}
\subsection{Problems and possible improvements}%
		\begin{itemize}
			\item The framework allows for online manual parameter tuning, however, it would be beneficial to include an auto-tuner to facilitate performance matching
			\item Code generation is beneficial for the implementation of rapid prototyping such as in this project, nevertheless, it often results in very large source codes that are hard to debug rendering the detailed function profiling more complex and one could just avoid code generation
		\end{itemize}

	\section{Conclusion}
	This paper presents a development framework for designing, validating, and implementing a real-time optimal controller for autonomous driving applications. The validation process satisfies the automotive industry requirements as it progresses from ISO standards scenarios, to the different XiL applications. The framework allows testing in real and/or virtual environments using high-fidelity dynamics and sensors in Simcenter software.
	The NMPC approach is demonstrated with the case of trajectory control as it is deployed as a low-level controller in the real-time applications at 25Hz, showing the potential capabilities of this controller type for collision and accident avoidance and ACC. \textcolor{black}{Trajectory control is one relevant use case of the applicability of real-time NMPC, however, other use cases could be tested using the same XiL development framework.} Embedded numerical optimization is deployed on the platform and applications can be extended to more complex OCP or other optimization based formulations such as optimal planning. The framework was tested and validated on a Ford Focus in the parking area and on proving grounds, in presence of obstacles and with a strict requirement on real-time calculations.

\bibliographystyle{ieeetr}
\bibliography{Biblio}

\end{document}